\documentclass[sigconf]{acmart}

\usepackage{algorithm}
\usepackage{algpseudocode}
\newcommand{\algname}[1] {{\fontfamily{cmtt}\selectfont {#1}}}
\usepackage{dsfont}

\usepackage{caption}
\usepackage{subcaption}
\usepackage{multirow}
\usepackage{graphicx,wrapfig,lipsum}

\usepackage{float}
\DeclareMathOperator{\EX}{\mathbb{E}}
\DeclareMathOperator*{\argmax}{argmax} 
\newcommand\norm[1]{\left\lVert#1\right\rVert}
\usepackage{bbm}
\newtheorem{theorem}{Theorem}
\usepackage{array}
\usepackage{xspace}
\newcommand{\header}[1]{\vspace*{1mm}\noindent\textbf{#1}.}
\usepackage{enumitem}

\AtBeginDocument{%
  \providecommand\BibTeX{{%
    \normalfont B\kern-0.5em{\scshape i\kern-0.25em b}\kern-0.8em\TeX}}}

\copyrightyear{2024}
\acmYear{2024}
\setcopyright{rightsretained}
\acmConference[CIKM '24]{Proceedings of the 33rd ACM International Conference on Information and Knowledge Management}{October 21--25, 2024}{Boise, ID, USA}
\acmBooktitle{Proceedings of the 33rd ACM International Conference on Information and Knowledge Management (CIKM '24), October 21--25, 2024, Boise, ID, USA}
\acmDOI{10.1145/3627673.3679763}
\acmISBN{979-8-4007-0436-9/24/10}

\begin{document}

\title{Mitigating Exposure Bias in Online Learning to Rank Recommendation: A Novel Reward Model for Cascading Bandits}

\author{Masoud Mansoury}
\orcid{0000-0002-9938-0212}
\affiliation{%
\institution{Delft University of Technology}
  \city{Delft}
  \country{The Netherlands}
}
\email{m.mansoury@tudelft.nl}
\authornote{Work done while the author was with Elsevier Discovery Lab and University of Amsterdam.}

\author{Bamshad Mobasher}
\orcid{0000-0001-9701-9178}
\affiliation{%
\institution{DePaul University}
  \city{Chicago}
  \country{USA}
}
\email{mobasher@cs.depaul.edu}

\author{Herke van Hoof}
\orcid{0000-0002-1583-3692}
\affiliation{%
\institution{University of Amsterdam}
  \city{Amsterdam}
  \country{The Netherlands}
}
\email{h.c.vanhoof@uva.nl}

\begin{abstract}
  Exposure bias is a well-known issue in recommender systems where items and suppliers are not equally represented in the recommendation results.  
  This bias becomes particularly problematic over time as a few items are repeatedly over-represented in recommendation lists, leading to a \textit{feedback loop} that further amplifies this bias.  
  Although extensive research has addressed this issue in model-based or neighborhood-based recommendation algorithms, less attention has been paid to online recommendation models, such as those based on top-$K$ contextual bandits, where recommendation models are dynamically updated with ongoing user feedback. In this paper, we study exposure bias in a class of well-known contextual bandit algorithms known as \textit{Linear Cascading Bandits}. We analyze these algorithms in their ability to handle exposure bias and provide a fair representation of items in the recommendation results. 
  Our analysis reveals that these algorithms fail to mitigate exposure bias in the long run during the course of ongoing user interactions. We propose an \textbf{E}xposure-\textbf{A}ware reward model that updates the model parameters based on two factors: 1) implicit user feedback and 2) the position of the item in the recommendation list. The proposed model mitigates exposure bias by controlling the utility assigned to the items based on their exposure in the recommendation list. Our experiments with two real-world datasets show that our proposed reward model improves the exposure fairness of the linear cascading bandits over time while maintaining the recommendation accuracy. It also outperforms the current baselines. Finally, we prove a high probability upper regret bound for our proposed model, providing theoretical guarantees for its performance.
\end{abstract}

\ccsdesc[500]{Information systems~Users and interactive retrieval}
\ccsdesc[500]{Information systems~Recommender systems}

\keywords{recommender systems, contextual bandits, exposure fairness}

\maketitle

\section{Introduction}

Recommender systems utilize users’ interaction data on different items to generate personalized recommendations \cite{resnick1994grouplens,aggarwal2016recommender,burke2011recommender,lu2012recommender}. Traditionally, the success of these systems are measured based on the degree to which the recommendations generated matches the user preferences \cite{shani2011evaluating,herlocker2004evaluating}. However, this user-centric view for building recommendation models, while captures users' preferences, neglects the item-side utilities, or what is referred to as \textit{exposure bias} \cite{mansoury2021understanding,singh2018fairness,heuss2023predictive}.

\header{Problem definition} 
Exposure bias in recommender systems refers to the phenomenon of items not being uniformly represented in the recommendation results: Few items are frequently shown in the recommendation lists, while the majority of other items rarely appear in the recommendation results~\cite{mansoury2021understanding,chen2023bias}. 
This bias, if not addressed, can result in a number of negative consequences for system performance. First, it can impact economic gains for items for suppliers of underexposed items, leading to unfair treatment and disincentivizing participation in the marketplace~\cite{mehrotra2018towards}. Secondly, it may hinder the system's ability to provide useful, but less popular, recommendations to consumers \cite{ciampaglia2018algorithmic,huang2024going}. Finally, users have a greater chance of interacting with over-exposed items, perpetuating their prominence in future recommendations, and amplifying existing biases. Amplification of exposure bias for a few items would be at the expense of the under-exposure for a majority of other items (including some that might be 
 of interest for some users) and consequently may push those items out of the marketplace\cite{mansoury2020feedback,sinha2016deconvolving,mansoury2023fairness}. 

\header{Research gap}
Most existing research to study exposure bias has focused on classical recommendation models in static settings where a single round of recommendation results is analyzed \cite{mansoury2021understanding,suhr2019two}. Although these studies reveal important aspects of exposure bias and propose solutions to tackle it, the long-term impact of this bias on online learning-to-rank recommendation models has yet to be explored significantly. This is a research gap which we seek to remedy in this paper. Filling this gap requires studying the task of recommendation problem in dynamic and interactive settings where users are engaged in ongoing interaction with the system and preference models are dynamically updated over time. 

In our study, we focus specifically on \textit{Cascading Bandits} (CB) \cite{kveton2015cascading,kveton2015combinatorial,li2016contextual,zong2016cascading} which provide a principled solution for online learning of recommendation models. The ability of CB to handle \textit{ position bias} \cite{collins2018position,hofmann2014effects} and perform \textit{ exploration} \cite{cao2024does,barraza2017exploration,mcinerney2018explore} makes it an interesting choice for developing online recommendation algorithms. The main question in this research is how CB distribute exposure among items in the system? Although these algorithms perform exploration in the items space to collect user feedback on different items, our study in this paper shows that this exploration does not necessarily lead to a sufficiently fair exposure for items in the long run. 

\header{Contributions and findings}
In this paper, we study exposure bias in cascading bandits and introduce a novel reward model to mitigate exposure bias in these algorithms. In cascading bandit algorithms, all selected items in the recommendation lists are similarly rewarded, regardless of what their position is in the list. This means that a clicked item on top of the list (e.g., at the first position) is equally rewarded as a clicked item at the bottom of the list (e.g., at position K). Also, the same formulation is considered to penalize ignored (unclicked) items. We hypothesize that considering the positional information of clicked/unclicked items when rewarding/penalizing those items would not only lead to a better adaptation of the model to user feedback, but also, most importantly, lead to a significant reduction in exposure bias over time.

We propose an \textbf{E}xposure-\textbf{A}ware (EA) reward model and integrate it into the existing cascading bandit algorithms. Our reward model updates the bandit model parameters based on two factors: 1) the user feedback on recommended items, whether the item is clicked or not, and 2) the position of the item in the list. In fact, the proposed model rewards or penalizes the clicked or unclicked items, respectively, based on their position in the recommendation list. 
This control over the degree of reward or penalization for items based on their exposure in the recommendation lists incentivizes more exploration and reduces exposure bias on items. Extensive experiments on two real-world datasets show that the proposed reward model not only reduces the exposure bias in cascading bandits, but also outperforms the state-of-the-art baselines in mitigating exposure bias while maintaining the recommendation accuracy. We also show theoretical guarantees for the performance of our reward model by proving a high probability upper regret bound for it.

\section{Background}\label{background}

In this section, we review the CB and the definitions of exposure fairness in recommender systems. Formally, $\mathcal{I}=\{1,...,m\}$ be the set of all items in the system. The task of generating recommendations in each round $t \in \{1,2,...,n\}$ is delivering a recommendation list of size $K$ to a target user. Let denote this recommendation list as $\mathcal{L}_t \in \Pi_K(\mathcal{I})$, where $\Pi_K(\mathcal{I})$ is the set of all $K$-permutations of the set $\mathcal{I}$. $\mathcal{L}(k)$ denotes the item in the $k$-th position of $\mathcal{L}$.

\subsection{Cascading bandit} \label{cb}

The Cascade Bandit (CB) is a principled method of operationalizing recommendation models in an online environment under the assumption that users will behave according to a cascade model~\cite{kveton2015cascading,zong2016cascading}. The cascade click model \cite{craswell2008experimental} is a well-known click model to interpret the click behavior of users on the recommendation list. Given the recommendation list $\mathcal{L}$, the target user examines each recommended item in $\mathcal{L}$ from the first position to the last, clicks on the first attractive item and stops examining the rest of the items. In this way, the items above the clicked item are considered unattractive, the clicked item is considered attractive, and the rest of the items are considered as unobserved. The probability that a user clicks on an item $\mathcal{L}(k)$ is called \textit{attraction probability} and we denote it as $\omega(\mathcal{L}(k))$. 
In the following, we describe the cascading bandit formulation for a user $u$ interacting with the system.  

Cascading bandits can be represented by a tuple ($\mathcal{I}$,$K$,$P$), where $P$ is a probability distribution over a binary hypercube $\{0,1\}^\mathcal{I}$. Also, let $\text{w}_t \in \{0,1\}^\mathcal{I}$ denote the preference weights for each item drawn from $P$, the degree to which $u$ is interested to each item where $\text{w}_t(\mathcal{L}(i))=1$ signifies that the item $\mathcal{L}(i)$ attracts $u$ in round $t$. Also, assuming that the preference weights of items in the ground set $\mathcal{I}$ are independently distributed as:
\begin{equation}
    P(\text{w}) = \prod_{i \in \mathcal{I}}{\text{Ber}_{\omega(i)}(\text{w}(i))}
\end{equation}

\noindent where $\text{Ber}_{\omega(i)}(.)$ is the Bernoulli distribution with mean $\omega(i)$. 

In each round $t$, the learning agent provides a recommendation list of size $K$, $\mathcal{L}_t \in \Pi_K(\mathcal{I})$, to the target user. According to the cascade click model, the user examines $\mathcal{L}_t$ from the first item (i.e., $\mathcal{L}(1)$) to the last one (i.e., $\mathcal{L}(K)$) and clicks on the first item of interest. We use $C_t \in \{1,..,K,K+1\}$ to denote the position of the clicked item. Note that $C_t \leq K$ holds if user clicks on an item in $\mathcal{L}_t$, otherwise $C_t=K+1$. Since user only clicks on the first "attractive" item, $\text{w}_t(\mathcal{L}(k))$ can be defined as: 
\begin{equation}\label{cascade_weight}
    \text{w}_t(\mathcal{L}(k))=\mathbbm{1}(C_t=k), \;\;\;\;\text{where}\;\; k \in [1,...,\min\{K,C_t\}]
\end{equation}

\noindent where $\mathbbm{1}[.]$ is the indicator function returning zero when its argument is False and 1 otherwise. And the reward is defined as:
\begin{equation}\label{eq_reward}
    \mathcal{R}(\mathcal{L}_t,\text{w}_t) = 1 - \prod_{i=1}^{K}{(1-\text{w}_t(\mathcal{L}_t(i)))}
\end{equation}

The goal of the agent is to minimize the disparity in reward observed on the generated recommendation list by the agent and the optimal ranker (or equivalently maximizing the number of clicks observed on recommended items) and can be computed as:
\begin{equation}\label{regret}
    R(n)=\EX{\left[\sum_{t=1}^{n}{\mathcal{R}(\mathcal{L}^*,\text{w}_t)-\mathcal{R}(\mathcal{L}_t,\text{w}_t)}\right]}
\end{equation}

\noindent where $\mathcal{L}^*$ is the \textit{optimal recommendation list} that maximizes the reward at each time $t$ and is computed as follows.
\begin{equation}
    \mathcal{L}^*=\argmax_{\mathcal{L} \in \prod_{K}{(\mathcal{I})}}{\mathcal{R}(\mathcal{L},\omega)}
\end{equation}

\subsection{Measuring exposure fairness}\label{sec_exp_fairness_measure}

In the existing literature, there are many metrics available to measure exposure fairness in recommender systems \cite{raj2022measuring,singh2018fairness,zehlike2020reducing,heuss2022fairness,liu2024measuring}. In this study, our focus is on assessing exposure fairness through various dimensions within the family of exposure metrics. Specifically, we scrutinize two critical dimensions: (i) consideration or disregard of the item's position in the recommendation list (w/ or w/o position, respectively), and (ii) allocation of exposure proportionately or irrespective of items' merit (w/ or w/o merit, respectively). Table \ref{tbl_fair_metrics} provides an overview of four distinct notions of exposure based on these dimensions.\footnote{Merit, in this context, refers to any quality measure for items, such as relevance \cite{balagopalan2023role}. The definition of the merit measure employed in this paper is elucidated in Section \ref{experiment}.}

\begin{table}[t]
\caption{Four different notions of exposure for items ($\mathcal{U}$ is the set of all users, $\mathcal{L}_{u}$ is the recommendation list delivered to user $u$, and $K$ is the size of the recommendation list).}
\vspace{-15pt}
\label{tbl_fair_metrics}
\centering
\setlength{\tabcolsep}{2pt}
\renewcommand{\arraystretch}{1.5}
    \begin{tabular}{m{0.1in}| p{1.6in}| p{1.55in}|}
    \multicolumn{1}{r}{}
     & \multicolumn{1}{c}{\textbf{w/o merit}}
     & \multicolumn{1}{c}{\textbf{w/ merit}} \\
    \cline{2-3}
    \multirow{4}[1]{*}{\rotatebox[origin=c]{90}{\textbf{w/o position}}} & \centering\arraybackslash\textit{Exposure is binary without considering item's merit \newline\newline $E^B(i)= \sum\limits_{u \in \mathcal{U}}{\mathbbm{1}[i \in \mathcal{L}_{u}]}$\newline} & \centering\arraybackslash\textit{Exposure is binary in proportion to item's merit \newline\newline $E^{BM}(i)=\frac{E^B(i)}{merit(i)}$\newline} \\
    \cline{2-3}
    \multirow{4}[1]{*}{\rotatebox[origin=c]{90}{\textbf{w/ position}}} & \centering\arraybackslash\textit{Exposure depends on the position without considering item's merit \newline\newline \text{\footnotesize $E^P(i)= \sum\limits_{u \in \mathcal{U}}\sum\limits_{k=1}^K{\mathbbm{1}[i \in \mathcal{L}_{u}]\frac{1}{\log_2(1+k)}}$}\newline} & \centering\arraybackslash\textit{Exposure depends on position in proportion to item's merit \newline\newline $E^{PM}(i)=\frac{E^P(i)}{merit(i)}$\newline} \\
    \cline{2-3}
\end{tabular}
\end{table}

Exposure fairness, as we define it, refers to the equitable distribution of exposure among items. With an exposure distribution representing the allocated exposure value for each item, our objective is to assess the extent to which this distribution achieves uniformity, with a uniform distribution being deemed the fairest. The Gini index \cite{antikacioglu2017post,vargas2014improving} is a well-known metric to measure the uniformity of a distribution. Given that the Gini index falls within the range of $[0,1]$, for consistency, we report 1 minus the Gini index in this paper. Consequently, a Gini index value of 1 signifies the fairest outcome, while a value of 0 denotes the most unfair outcome. Calculating the Gini index on the exposure distribution derived from each definition outlined in Table \ref{tbl_fair_metrics} yields four notions of exposure fairness:

\begin{itemize}[leftmargin=17pt]
    \item \textbf{Equality of binary exposure (Equality$^{(B)}$)}: This computes the Gini Index over the exposure distribution of $E^B$ (i.e., w/o position and w/o merit). 
    \item \textbf{Equality of position-based exposure (Equality$^{(P)}$)}: This computes the Gini Index over the exposure distribution of $E^P$ (i.e., w/ position and w/o merit).
    \item \textbf{Equity of binary exposure (Equity$^{(B)}$)}: This computes the Gini Index over the exposure distribution of $E^{BM}$ (i.e., w/o position and w/ merit).
    \item \textbf{Equity of position-based exposure (Equity$^{({P})}$)}: This computes the Gini Index over the exposure distribution of $E^{PM}$ (i.e., w/ position and w/ merit).
\end{itemize}

\section{Exposure-Aware Cascading Bandits}\label{section_eacb}

The previous approaches based on cascading bandits do not consider the position of clicked or ignored (unclicked) items when assigning rewards or penalties\footnote{For the rest, when an item is examined, but it is not clicked, we call it an "unclicked" item. This is different from the unobserved items that have not even been examined.}. This means that items clicked on at the top of the list receive the same reward as those clicked at the bottom. Users tend to select highly exposed items, often positioned at the top, either due to their accessibility or genuine interest~\cite{granka2008eyetracking,lorigo2008eye}. Conversely, less exposed items towards the bottom of the list require more effort from users to discover, and when clicked, are likely of higher importance. Thus, to better adjust the model to user behavior, clicked items at the bottom should be rewarded more than those at the top. This also provides additional incentives for assigning more exposure to less exposed items in the future.

Similarly, unclicked items should be penalized differently depending on their position. Unclicked items on the top should receive a greater penalty than those at the bottom. When a highly exposed item is not clicked, it suggests that the recommendation model inaccurately assumed that it was of high interest. Penalizing such items more heavily than less exposed unclicked items helps refine the model to avoid prioritizing them in future recommendations. This also incentives for downgrading recently over exposed items and promoting less exposed items in the future. To address these issues, we propose an Exposure-Aware Cascading Bandit (EACB) that adjusts rewards based on the position of clicked items in the recommendation list. Hence, we reformulate Eq. \ref{cascade_weight} as:
\begin{equation}\label{ea_weighting}
    \text{w}^{EA}_t(\mathcal{L}(k))=\mathcal{F}_{t,k}\times \mathbbm{1}[C_t=k] - \gamma\mathcal{F}_{t,k}\times \mathbbm{1}[C_t<k]
\end{equation}

\begin{table}[t]
\caption{Weighting functions to define $\mathcal{F}_{t,k}$ in Eq. \ref{ea_weighting} ($k$ is the position of the item in the list and $t$ is the current round).}
\label{tbl_weight_funcs}
\vspace{-10pt}
\centering
\begin{tabular}{@{}llll@{}}
\toprule
\textbf{Function}                & \textbf{Abbreviation} & \textbf{Formula} & \textbf{Parameters} 
\\ 
\toprule
Logarithmic     & Log & $\log(1+k)$      & $-$     
\\
\midrule
Exponential & RBP     &  $\beta^{k-1}$       & patience $\beta$   
\\
\midrule
Linear     & Linear & $\beta \times k$         & patience $\beta$ 
\\ 
\bottomrule
\end{tabular}

\end{table}

\begin{figure}
  \centering
  \includegraphics[width=\linewidth]{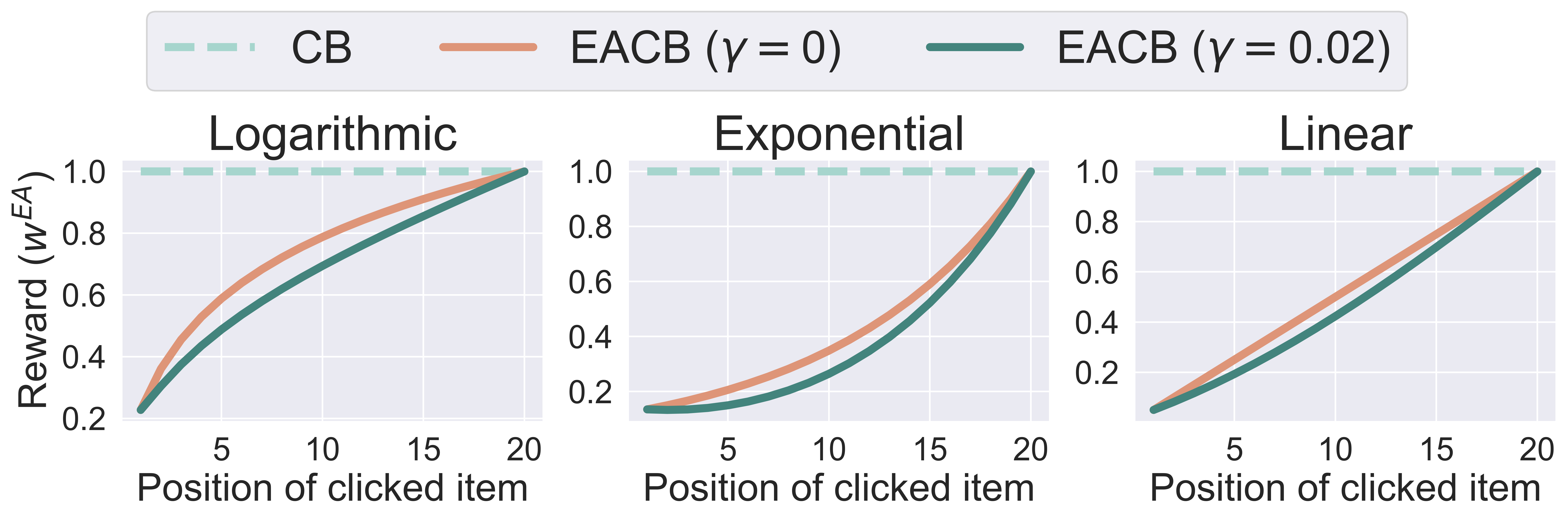}
  \vspace{-22pt}
  \caption{Reward distribution of CB and EACB for different weight functions when click is observed at varying positions in the list.}
    \label{fig:rewards}
\end{figure}

\noindent where $k \in [1,...,\min\{K,C_t\}]$ and $\mathcal{F}_{t,k}$ is the \textit{exposure-aware weight function} that assigns weights to all examined items based on their position in the recommendation list. The indicator function ensures that the appropriate term (reward or penalty) is applied based on whether the item is clicked or unclicked: if the examined item is clicked, the first term (reward) applies as $\mathbbm{1}[C_t=k]=1$ and $\mathbbm{1}[C_t<k]=0$, otherwise, the second term (penalization) applies as $\mathbbm{1}[C_t=k]=0$ and $\mathbbm{1}[C_t<k]=1$. The hyperparameter $\gamma$ controls the degree of penalization for unclicked items. A small $\gamma$ value allows for slight penalization of unclicked items, with the focus primarily on learning user preferences from clicked items.

The choice of $\mathcal{F}_{t,k}$ is crucial for an effective exposure-aware reward model. It must meet two criteria: 1) positively weight clicked items at the bottom more than those at the top, and 2) negatively weight unclicked items at the top more than those at the bottom. In this paper, we consider three different weight functions, outlined in Table \ref{tbl_weight_funcs}, which align with established browsing models~\cite{srikant2010user}. For example, the logarithmic function follows the standard exposure drop-off~\cite{singh2018fairness} used in ranking metrics (e.g., nDCG), while the exponential function follows Rank-Biased Precision (RBP)~\cite{moffat2008rank}. Figure~\ref{fig:rewards} shows the reward distribution of CB and EACB for different weight functions, showcasing the varying intensities of weighting assigned to observed clicked items. The logarithmic function exhibits the highest intensity, followed by the linear and exponential functions.

According to the cascade click model and Eq. \ref{ea_weighting}, the examination probability of an item $\mathcal{L}(k)$ in EACB would be:
\begin{equation}
    \prod_{i=1}^{k-1}{(1-(\mathcal{F}_{t,k}\times \mathbbm{1}[C_t=k] - \gamma\mathcal{F}_{t,k}\times \mathbbm{1}[C_t<k]))}
\end{equation}

\noindent and the expectation of reward at round $t$ is computed as follows:
\begin{equation}\label{eq_eareward}
    \mathcal{R}(\mathcal{L}_t,\text{w}_t^{EA}) = 1 - \prod_{i=1}^{K}{(1-(\mathcal{F}_{t,k}\times \mathbbm{1}[C_t=k] - \gamma\mathcal{F}_{t,k}\times \mathbbm{1}[C_t<k]))}
\end{equation}

\subsection{Algorithm for learning EACB}\label{section_alg}

Various algorithms have been developed within the Cascading Bandit (CB) framework \cite{zong2016cascading,hiranandani2020cascading,li2020cascading}. In this paper, we specifically concentrate on the linear cascading bandit proposed by Zong et al. \cite{zong2016cascading} and extend it to incorporate our exposure-aware reward model.

The algorithm \ref{alg_eacm} presents the algorithmic process of our EACB. In each round, the algorithm computes the \textit{attraction probability}, $w(i)$, of a target item $i$, representing the likelihood of the target user liking the item. This probability is derived from the dot product of the item features, $x_i$, and the user preference vector $\theta^*$, denoted as $w(i)=\theta^*x_i^T$. While item features are known to the algorithm, the user preference vector $\theta^*$ is unknown and must be learned through user interactions. Thus, in the initial step (line 4), the algorithm estimates the user preference vector from past observations on item features and their corresponding attraction probabilities.

This estimation process can be framed as a ridge regression problem, where $\hat{\theta}_t$ is computed as:
\begin{equation}
    \hat{\theta}_t= (X_t^TX_t+\lambda I)^{-1}X_t^T \hat{W}_t
\end{equation}

\noindent where $X_t\in \mathbb{R}^{m\times d}$ is the matrix of item features, $\hat{W}_t\in \mathbb{R}^{m\times 1}$ is the vector of items' attraction probabilities at round $t$, and $\lambda$ is the regularization term. The algorithm iteratively updates the model parameters $M_t=X_t^TX_t+\lambda I$ and $B_t=X_t^T \hat{W}_t$.

To address uncertainty in estimating user preferences and enable exploration in the item space, an item selection strategy is employed. Examples of these strategies are $\epsilon$-Greedy \cite{sutton2018reinforcement}, Upper Confidence Bound \cite{auer2000using,lai1985asymptotically}, and Thompson Sampling \cite{thompson1933likelihood,chapelle2011empirical}. In this paper, we focus on the Upper Confidence Bound (UCB) item selection strategy and leave the investigation on other strategies as our future work. According to UCB, the score for each item $i$ is predicted by combining the estimation of attraction probability with an upper bound (line 7), as expressed by:
\begin{equation}\label{eq_ucb}
    U_t(i)=\hat{\theta}_{t}x_i^T+\alpha\sqrt{x_iM_{t-1}^{-1}x_i^T}
\end{equation}

\noindent where $M_{t-1}\in \mathbb{R}^{d\times d}$ is the co-variance matrix of item features. The term $\sqrt{x_iM_{t-1}^{-1}x_i^T}$ is the upper bound for the estimated weight of item $i$ which covers the optimal weight and is computed by norm of $x_i$ weighted by $M_{t-1}^{-1}$ (i.e., $\norm{x_i}_{M_{t-1}^{-1}}$). $\alpha$ is a hyperparameter that controls the degree of exploration. Given scores computed for each item using Eq. \ref{eq_ucb}, $K$ items with the largest $U_t(i)$ are returned as the recommendation list (lines 9-12).

Upon receiving feedback from the user for the recommendation list $\mathcal{L}_t$ (line 14), the agent updates the model parameters (lines 16-23) based on the user's feedback. Specifically, if an examined item is clicked, the parameter $B_t$ is rewarded; otherwise, it is penalized.

It should be noted that our proposed EACB algorithm involves several tunable hyperparameters, including $\alpha$ for exploration control, $\sigma$ for the growth rate of $M_t$, $\lambda$ for regularization and $\gamma$ for the degree of penalization on unclicked items. Adjusting these hyperparameters enables fine-tuning of the algorithm's performance.

\begin{algorithm}[t!]
    \caption{Exposure-aware cascading bandit algorithm}\label{alg_eacm}
    \begin{algorithmic}[1]
        \Require Number of rounds $n$, size of recommendation list $K$, number of feature $d$, learning rate $\sigma$
        \State \textcolor{purple}{// Initialization}
        \State \textbf{M}$\leftarrow\lambda I^{d\times d}$, \textbf{B} $\leftarrow\text{0}^{d}$
        \For{$t=1,...,n$}
            \State $\hat{\theta}_t\leftarrow \sigma^{-2}M_{t-1}^{-1}B_{t-1}$
            \State \textcolor{purple}{// Recommend a list of $K$ items}
            \For{$i \in \mathcal{I}$}
            \State Compute $U_t(i)$ using Eq. \ref{eq_ucb}
            \EndFor
            \For{$k=1,...K$}
                \State $\textbf{i}_k \leftarrow \argmax_{e \in \mathcal{I}\textbackslash\{\textbf{i}_1,...,\textbf{i}_{k-1}\}}{\mathcal{S}_t(e)}$
            \EndFor
            \State $\mathcal{L}_t \leftarrow (\textbf{i}_1,...,\textbf{i}_K)$
            \State \textcolor{purple}{// Collect user's feedback on $\mathcal{L}$}
            \State Display $\mathcal{L}_t$ and observe click feedback $C_t \in \{1,...,K,K+1\}$
            \State \textcolor{purple}{//Update model parameters}
            \For{$k=1,...,\min\{K,C_t\}$}
                \State $M_t\leftarrow M_{t-1}+\sigma^{-2} x_{\mathcal{L}(k)}^Tx_{\mathcal{L}(k)}$
                \If{$C_t==k$}
                    \State $B_t \leftarrow B_{t-1} + \mathcal{F}_{t,k}x_{\mathcal{L}(k)}$
                \Else
                    \State $B_t \leftarrow B_{t-1} - \gamma\mathcal{F}_{t,k}x_{\mathcal{L}(k)}$
                \EndIf
            \EndFor
        \EndFor
    \end{algorithmic}
\end{algorithm}

\section{Analysis of regret upper-bound}

In this section, we present the upper bound of $n$-step-regret for our proposed exposure-aware cascading bandits. Our analysis shows that with an extra condition (the choice of $\gamma$), our exposure-aware cascading bandit has the same upper bound for $n$-step-regret as the original cascading bandits \cite{zong2016cascading} as follows:
\begin{theorem}\label{theorem}
For any $\sigma>0$, $\norm{\theta^*}_2\leq1$, and
\begin{equation}\label{eq_alpha}
    \alpha \geq \frac{1}{\sigma}\sqrt{d \log{\left(1+\frac{nK}{d\sigma^2}\right)}+2\log{(n)}}+\norm{\theta^*}_2
\end{equation}
\begin{equation}\label{eq_gamma}
    \gamma \leq \frac{1}{\mathcal{F}_{t,k}}-1 \;,\;\;\;\;\forall k \in \{1,...,K\}
\end{equation}
\noindent we have,
\begin{equation}\label{eq_regret_bound_theorem}
    R(n) \leq 2 \alpha K\sqrt{\frac{dn\log{\left[1+\frac{nK}{d\sigma^2}\right]}}{\log{\left(1+\frac{1}{\sigma^2}\right)}}} + 1.
\end{equation}
\end{theorem}

This theorem implies that for sufficiently optimistic $\alpha$ and $\gamma$, combining the equations \ref{eq_alpha}, \ref{eq_gamma}, and \ref{eq_regret_bound_theorem}, we have $R(n)=\mathcal{O}(dK\sqrt{n})$ where $\mathcal{O}$ ignores logarithmic factors. This is also the same upper bound for the original linear cascading bandits \cite{zong2016cascading}. Moreover, this bound signifies two properties: (1) it states a near optimal bound with factor $\sqrt{n}$, (2) the bound is linear in the size of the recommendation lists and the number of features, which is a common dependence in learning bandit algorithms \cite{abbasi2011improved}. 

\subsection{Proof of Theorem 1}\label{theorem_proof}

Let $\Pi(\mathcal{I})=\bigcup_{i=1}^{m}{\Pi_i(\mathcal{L})}$ be all possible recommendation lists in the item catalog $\mathcal{I}$ and $O:\Pi(\mathcal{I})\leftarrow[0,1]$ be an arbitrary weight function for the lists. According to the reward model in Eq. \ref{eq_eareward}, the expected reward of a recommendation list can be computed as:
\begin{equation}\label{eq_expected_reward}
    f(\mathcal{L},O)= 1-\prod_{i=1}^K\left(1-\mathcal{F}_{t,i}\times O(\mathcal{L}(i)) + \gamma\times\mathcal{F}_{t,i}\times (1-O(\mathcal{L}(i))\right)
\end{equation}

For each item in $\mathcal{L}$ we define $O$, its high probability upper-bound $H_t$, and its high probability lower-bound $L_t$ as:
\begin{equation}\label{eq_higher_lower_bound}
\begin{split}
    &O(\mathcal{L}(i)) = \theta^*x_{\mathcal{L}(i)}^{T}
    \\
    &H_t(\mathcal{L}(i)) = \text{Func}_{[0,1]}\left(\hat{\theta}_tx_{\mathcal{L}(i)}^{T} + \alpha \sqrt{x_{\mathcal{L}(i)}M_t^{-1}x_{\mathcal{L}(i)}^T}\right)
    \\
    &L_t(\mathcal{L}(i)) = \text{Func}_{[0,1]}\left(\hat{\theta}_tx_{\mathcal{L}(i)}^{T} - \alpha \sqrt{x_{\mathcal{L}(i)}M_t^{-1}x_{\mathcal{L}(i)}^{T}}\right)
\end{split}
\end{equation}

\noindent where $\text{Func}_{[0,1]}(.)=max(0,min(1,0))$, projecting the estimated value onto range $[0,1]$. We also define the following notation: 
\begin{equation}
\begin{split}
    &\psi^\mathcal{X}_{t,i}=\mathcal{F}_{t,i}\times \mathcal{X}(\mathcal{L}(i)), \;\;\; \psi^{\prime \mathcal{X}}_{t,i}=\mathcal{F}_{t,i}\times (1-\mathcal{X}(\mathcal{L}(i))
\end{split}
\end{equation}

\noindent where $\mathcal{X}$ refers to one of the functions ($O$, $H_t$, or $L_t$) in Eq. \ref{eq_higher_lower_bound}. Now, we start our proof by defining event 
\begin{equation*}
    \mathcal{E}_t=\{L_t(\mathcal{L}(i))\leq O(\mathcal{L}(i)) \leq H_t(\mathcal{L}(i)), \forall i \in [1,K], \forall \mathcal{L}\in \Pi (\mathcal{I}) \}
\end{equation*}

\noindent and $\Bar{\mathcal{E}}_t$ as its complement. $\mathcal{E}_t$ contains all the lists that the attraction probability estimation of its items falls into the upper and lower confidence bound which is the main ingredient of the UCB item selection strategy. We derive the regret bound for a single time step $t$ and then extend it to the upper bound of cumulative regret of $n$ time steps. Hence, we have,
\begin{equation}
\begin{split}
    &\EX\left[\mathcal{R}(\mathcal{L}^*,\omega)-\mathcal{R}(\mathcal{L}_t,\text{w}_t)\right]=\EX\left[f(\mathcal{L}^*,O)-f(\mathcal{L}_t,O)\right]
    \\
    &\overset{a}\leq P(\mathcal{E}_t)\EX\left[f(\mathcal{L}^*,O)-f(\mathcal{L}_t,O)\right] + P(\Bar{\mathcal{E}}_t)
    \\
    &\overset{b}\leq P(\mathcal{E}_t)\EX\left[f(\mathcal{L}^*,H_t)-f(\mathcal{L}_t,O)\right] + P(\Bar{\mathcal{E}}_t)
    \\
    &\overset{c}\leq P(\mathcal{E}_t)\EX\left[f(\mathcal{L}_t,H_t)-f(\mathcal{L}_t,O)\right] + P(\Bar{\mathcal{E}}_t)
\end{split}
\end{equation}

\noindent where (a) holds because $\EX\left[f(\mathcal{L}^*,O)-f(\mathcal{L}_t,O)\right]\leq 1$; (b) holds because given the inequality
\begin{equation}\label{inequality_b}
    f(\mathcal{L}^*,H_t) \leq \max_{\mathcal{L}\in \prod_K{(\mathcal{I})}}{f(\mathcal{L},H_t)} \leq f(\mathcal{L}_t,H_t)
\end{equation}

\noindent and event $\mathcal{E}_t$, we have $f(\mathcal{L}^*,O) \leq f(\mathcal{L}^*,H_t)$; (c) holds because $f(\mathcal{L}^*,H_t)-f(\mathcal{L}_t,O) \leq \left[f(\mathcal{L}_t,H_t)-f(\mathcal{L}_t,O)\right]$ from Eq. \ref{inequality_b}.

Let $\mathcal{H}_t$ be the history of data collected up to time $t$. Then, for any $\mathcal{H}_t$ such that $\mathcal{E}_t$ holds, together with Eq. \ref{eq_expected_reward} and \ref{eq_higher_lower_bound}, we have,
\begin{equation*}
\begin{split}
    &\text{\footnotesize $f(\mathcal{L}_t,H_t)-f(\mathcal{L}_t,O)=\prod_{i=1}^K\left(1-\psi^O_{t,i} + (\gamma\times\psi^{\prime O}_{t,i})\right)-\prod_{i=1}^K\left(1-\psi_{t,i}^H + (\gamma\times\psi^{\prime H}_{t,i})\right)$}
    \\
    &\text{\footnotesize $\overset{a}{=}\sum_{i=1}^K{\left[\prod_{j=1}^{i-1}{\left(1-\psi^O_{t,j} + (\gamma\times\psi^{\prime O}_{t,j})\right)}\right]} \left(\psi_{t,i}^H - (\gamma\times\psi^{\prime H}_{t,i})-\psi_{t,i}^O + (\gamma\times\psi^{\prime O}_{t,i})\right)$}
    \\
    &\;\;\;\;\;\;\;\;\;\;\;\;\;\;\;\;\;\;\;\;\;\;\;\;\;\;\;\;\;\;\;\;\;\;\;\;\;\;\;\;\;\;\;\;\;\;\;\;\;\;\;\;\;\;\;\;\;\;\;\;\text{\footnotesize $\left[\prod_{k=i+1}^K{(1-\psi_{t,k}^H + (\gamma\times\psi^{\prime H}_{t,k}))}\right]$}
    \\
    &\text{\footnotesize $\overset{b}{\leq} \sum_{i=1}^K{\left[\prod_{j=1}^{i-1}{(1-\psi^O_{t,j} + (\gamma\times\psi^{\prime O}_{t,j}))}\right]} \left(\psi_{t,i}^H - (\gamma\times\psi^{\prime H}_{t,i})-\psi_{t,i}^O + (\gamma\times\psi^{\prime O}_{t,i})\right)$}
\end{split}
\end{equation*}

\noindent where (a) follows Lemma 1 in \cite{zong2016cascading}; and (b) holds because $\psi_{t,k}^H + \gamma\times\psi^{\prime H}_{t,k} \leq 1$. Now, we define the event $\mathcal{G}_{t,i}=\{\text{item }\mathcal{L}_t(i) \text{ is examined}\}$ where we have $\EX\left[\mathbbm{1}(\mathcal{G}_{t,i})\right]=\prod_{j=1}^{i-1}{(1-\psi_{t,j}^O + \gamma\times\psi^{\prime O}_{t,j})}$. Then, for any $\mathcal{H}_t$ under $\mathcal{E}_t$, we have,
\begin{equation*}
\begin{split}
    &\EX\left[f(\mathcal{L}_t,H_t)-f(\mathcal{L}_t,O)\;\middle|\;\mathcal{H}_t\right]
    \\
    & \leq \sum_{i=1}^K{\EX\left[\mathbbm{1}[\mathcal{G}_{t,i}]\;\middle|\;\mathcal{H}_t\right]\left(\psi_{t,i}^H - (\gamma\times\psi^{\prime H}_{t,i})-\psi_{t,i}^L + (\gamma\times\psi^{\prime L}_{t,i})\right)}
    \\
    & \overset{a}{\leq} 2\alpha \EX\left[\mathbbm{1}[\mathcal{G}_{t,i}]\sum_{i=1}^K{\left[\sqrt{x_{\mathcal{L}(i)}M_t^{-1}x_{\mathcal{L}(i)}^T}\right].\left[\mathcal{F}_{t,i}\times\left(1+\gamma\right)\right]}\;\middle|\;\mathcal{H}_t\right]
    \\
    & \overset{b}{\leq} 2\alpha \EX\left[\sum_{i=1}^{min\{K,C_t\}}{\left[\sqrt{x_{\mathcal{L}(i)}M_t^{-1}x_{\mathcal{L}(i)}^T}\right].\left[\mathcal{F}_{t,i}\times\left(1+\gamma\right)\right]}\;\middle|\;\mathcal{H}_t\right]
\end{split}
\end{equation*}

\noindent where (a) follows the definition of $H_t$ and $L_t$ from Eq. \ref{eq_higher_lower_bound}; and (b) follows the definition of $\mathcal{G}_{t,i}$. Thus, with $\phi_{t,\mathcal{L}(i)}=\sqrt{x_{\mathcal{L}(i)}M_t^{-1}x_{\mathcal{L}(i)}^T}$, the cumulative regret of $n$ rounds can be defined as:
\begin{equation}\label{eq_regret_bound}
\begin{split}
    &\text{\footnotesize $R(n) = \sum_{t=1}^n\EX\left[\mathcal{R}(\mathcal{L}^*,\omega)-\mathcal{R}(\mathcal{L}_t,\text{w}_t)\right]$}
    \\
    & \text{\footnotesize $\leq \sum_{t=1}^n \left[2\alpha \EX\left[\sum_{i=1}^{min\{K,C_t\}}{\phi_{t,\mathcal{L}(i)}\times\mathcal{F}_{t,i}\times\left(1+\gamma\right)}\;\middle|\;\mathcal{E}_t\right]P(\mathcal{E}_t)+P(\Bar{\mathcal{E}}_t)\right]$}
    \\
    & \text{\footnotesize $\leq2\alpha\EX\left[\sum_{t=1}^n{\sum_{i=1}^{min\{K,C_t\}}{\phi_{t,\mathcal{L}(i)}\times\mathcal{F}_{t,i}\times\left(1+\gamma\right)}}\right] + \sum_{t=1}^n{P(\Bar{\mathcal{E}}_t)}$}
\end{split}
\end{equation}

The regret bound can be derived by finding the worst-case bound on $\sum_{t=1}^n{\sum_{i=1}^{min\{K,C_t\}}{\left[\phi_{t,\mathcal{L}(i)}\times\mathcal{F}_{t,i}\times\left(1+\gamma\right)\right]}}$ and $\sum_{t=1}^n{P(\Bar{\mathcal{E}}_t)}$ terms in Eq. \ref{eq_regret_bound}. It should be noted that this is the same problem as in the original cascading bandits in \cite{zong2016cascading,hiranandani2020cascading,li2020cascading} except for the first term that contains an additional $\left[\mathcal{F}_{t,i}\times(1+\gamma)\right]$ term. \textbf{This is the main advantage of our proposed EACB which guarantees a lower upper-bound for the n-step-regret compared to CB. The reason is that with a proper choice for the value of $0 < \gamma < \frac{1}{\mathcal{F}_{t,i}}-1$, we have $\left[\mathcal{F}_{t,i}\times(1+\gamma)\right] < 1$ which leads to a smaller value for the first term (compared to CB):}
\begin{equation*}\label{ineq_simplify}
\begin{split}
    &\EX\left[\sum_{t=1}^n{\sum_{i=1}^{min\{K,C_t\}}{\phi_{t,\mathcal{L}(i)}\times\mathcal{F}_{t,i}\times\left(1+\gamma\right)}}\right]\leq \EX\left[\sum_{t=1}^n{\sum_{i=1}^{min\{K,C_t\}}{\phi_{t,\mathcal{L}(i)}}}\right]
\end{split}
\end{equation*}

\noindent where this inequality makes the first term in Eq. \ref{eq_regret_bound} similar to CB. Therefore, according to Lemma 2 in \cite{zong2016cascading}, we have,
\begin{equation}\label{eq_first_term}
    \sum_{t=1}^n{\sum_{i=1}^{min\{K,C_t\}}{\sqrt{x_{\mathcal{L}(i)}M_t^{-1}x_{\mathcal{L}(i)}^T}}} \leq K\sqrt{\frac{dn\log{\left[1+\frac{nK}{d\sigma^2}\right]}}{\log{\left(1+\frac{1}{\sigma^2}\right)}}}
\end{equation}

\noindent which is the worst-case bound for the first term in \ref{eq_regret_bound}. Also, for the second term in Eq. \ref{eq_regret_bound}, according to Lemma 3 in \cite{zong2016cascading}, we have $P(\Bar{\mathcal{E}}_t)\leq 1/n$ for any $\alpha$ that satisfies Eq. \ref{eq_alpha}. Therefore, together with Eq. \ref{eq_first_term} and \ref{eq_regret_bound}, we have,
\begin{equation}
    R(n) \leq 2 \alpha K\sqrt{\frac{dT\log{\left[1+\frac{TK}{d\sigma^2}\right]}}{\log{\left(1+\frac{1}{\sigma^2}\right)}}} + 1.
\end{equation}

\noindent which proves the Theorem \ref{theorem}.

\section{Experiments}\label{experiment}

Our experimental analysis on real-world datasets is designed to address the following research questions: (\textbf{RQ1}) What impact does adjusting the degree of exploration in the original linear cascading bandit have on exposure bias? (\textbf{RQ2}) Does our exposure-aware cascading bandit algorithm better mitigate the effect of exposure bias than existing exposure bias mitigation methods? (\textbf{RQ3}) How does varying the penalization parameter ($\gamma$) influence the performance of our exposure-aware cascading bandit algorithm?

\subsection{Datasets}
Our experiments are conducted on two publicly-available datasets: MovieLens 1M \cite{harper2015movielens} and Yahoo Music\footnote{R2- Yahoo! Music, \url{https://webscope.sandbox.yahoo.com/catalog.php?datatype=r}.}. The MovieLens dataset comprises 6K users who provided 1M ratings for 4K items. On the other hand, the Yahoo Music dataset contains ratings from 1.8M users for 136K songs, totaling 700M ratings. In both datasets, ratings fall within the range of $[1,5]$.

We follow the data preprocessing approach in \cite{hiranandani2020cascading,li2020cascading}. First, we map the ratings onto a binary scale: rating 4 and 5 are converted to 1 and other ratings to 0. Then, on MovieLens dataset, we create a sample of the data by extracting the 1000 most active users from the interaction data. On Yahoo Music dataset, we extract the 1000 most active users and the 1000 most rated songs from the interaction data. After this preprocessing, approximately 9\% and 1\% of the original ratings are retained for MovieLens and Yahoo Music, respectively.

\subsection{Evaluation metrics and baselines}

In our experimental evaluation, we investigate the impact of integrating our exposure-aware reward model into the linear cascading bandit algorithm \cite{zong2016cascading}. We compare the performance of this modified algorithm, termed \algname{EALinUCB}, with the original linear cascading bandit algorithm (\algname{LinUCB}), where our reward model is not employed. For brevity, we omit the term "Cascade" in the names of the algorithms. Additionally, we consider three variations of \algname{EALinUCB}, each utilizing a different weight function for training, as detailed in Table \ref{tbl_weight_funcs}. We also compare \algname{EALinUCB} with the following baselines:

\begin{itemize}[leftmargin=17pt]
    \item \textbf{Exposure-Aware aRm Selection (\algname{EARSLinUCB}) \cite{jeunen2021top}:} This method adopts a post-processing approach to improve exposure fairness. It reranks the generated recommendations by shuffling less relevant items to the bottom of the list, thus enhancing exposure fairness through randomization.
    \item \textbf{Fairness Regret Minimization (\algname{FRMLinUCB}) \cite{wang2021fairness}:} This baseline addresses exposure bias by formulating the bandit problem to minimize both reward regret and fairness regret. It assigns exposure to items proportionally to their merit, thereby promoting exposure fairness. Our implementation of fairness regret is based on the Equity$^{P}$ notion, where each item's exposure is proportional to its true relevancy score (see Section \ref{sec_exp_setup}). 
\end{itemize}

The key distinction between \algname{EARSLinUCB} and \algname{FRMLinUCB} compared to our proposed approach lies in their intervention strategy. Although these baselines intervene during the recommendation generation step, our exposure-aware cascading bandit algorithm intervenes during the reward/penalization step. This allows our approach to be more generalizable in various cascading bandit algorithms \cite{hiranandani2020cascading,li2020cascading}, as its effectiveness is not contingent on the performance of the underlying bandit algorithm. We leave the research on the generalizability of our EACB as our future work.

To measure the degree of exposure bias in each bandit algorithm, we utilize four metrics introduced in Section \ref{sec_exp_fairness_measure}. Higher values for these metrics indicate less exposure bias or a fairer exposure distribution among items. In addition, we evaluate the accuracy of the model using the following metrics:

\begin{itemize}[leftmargin=17pt]
    \item \textbf{Average number of clicks ($\overline{clicks}$):} This metric measures the total number of clicks (\#clicks) normalized by the number of users and number of rounds, providing insight into the model's performance in generating relevant recommendations:
    \begin{equation}
        \overline{clicks}=\frac{\#clicks}{|\mathcal{U}| \times n}
    \end{equation}
    \noindent where $0 \leq \overline{clicks} \leq 1$, 0 signifies no click and 1 indicates that all users clicked at least on one item at each round which is more desirable.
    \item \textbf{$n$-step-regret:} This metric measures cumulative regret, the difference in the observed number of clicks between the optimal ranker and the online ranker in $n$ rounds as defined in Eq. \ref{regret}. To ensure fair comparison, we use the original reward model (i.e., Eq. \ref{eq_reward}) for computing the $n$-step-regret for all algorithms, even though \algname{EALinUCB} employs a different reward model during the training process (i.e., Eq. \ref{eq_eareward}).
\end{itemize}

\begin{figure*}
  \begin{minipage}[b]{0.46\linewidth}
  \centering
    \includegraphics[width=\linewidth]{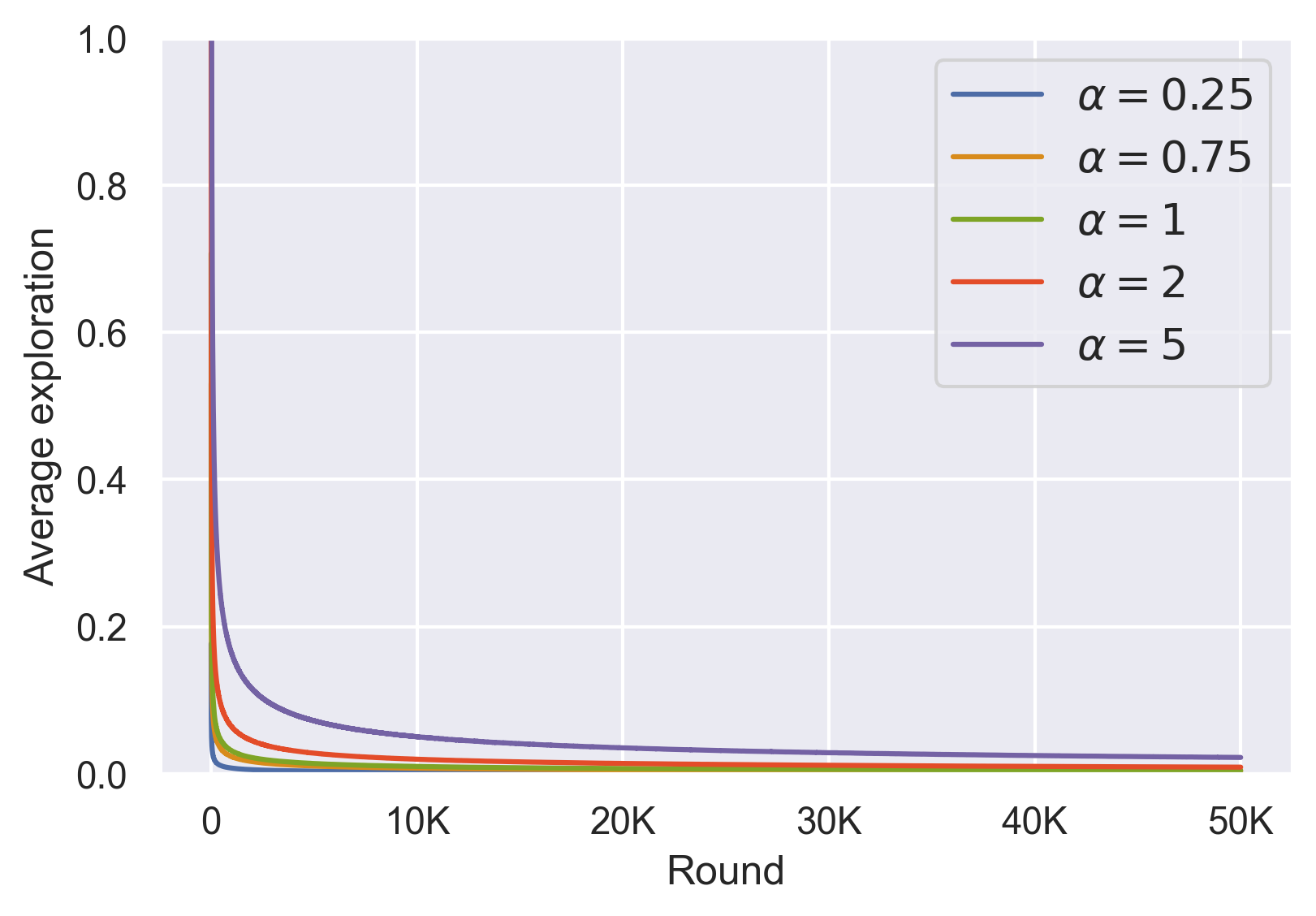}
  \end{minipage}\hfill
  \begin{minipage}[b]{0.53\linewidth}
  \centering
    \begin{subfigure}{0.33\linewidth}
      \includegraphics[width=\linewidth]{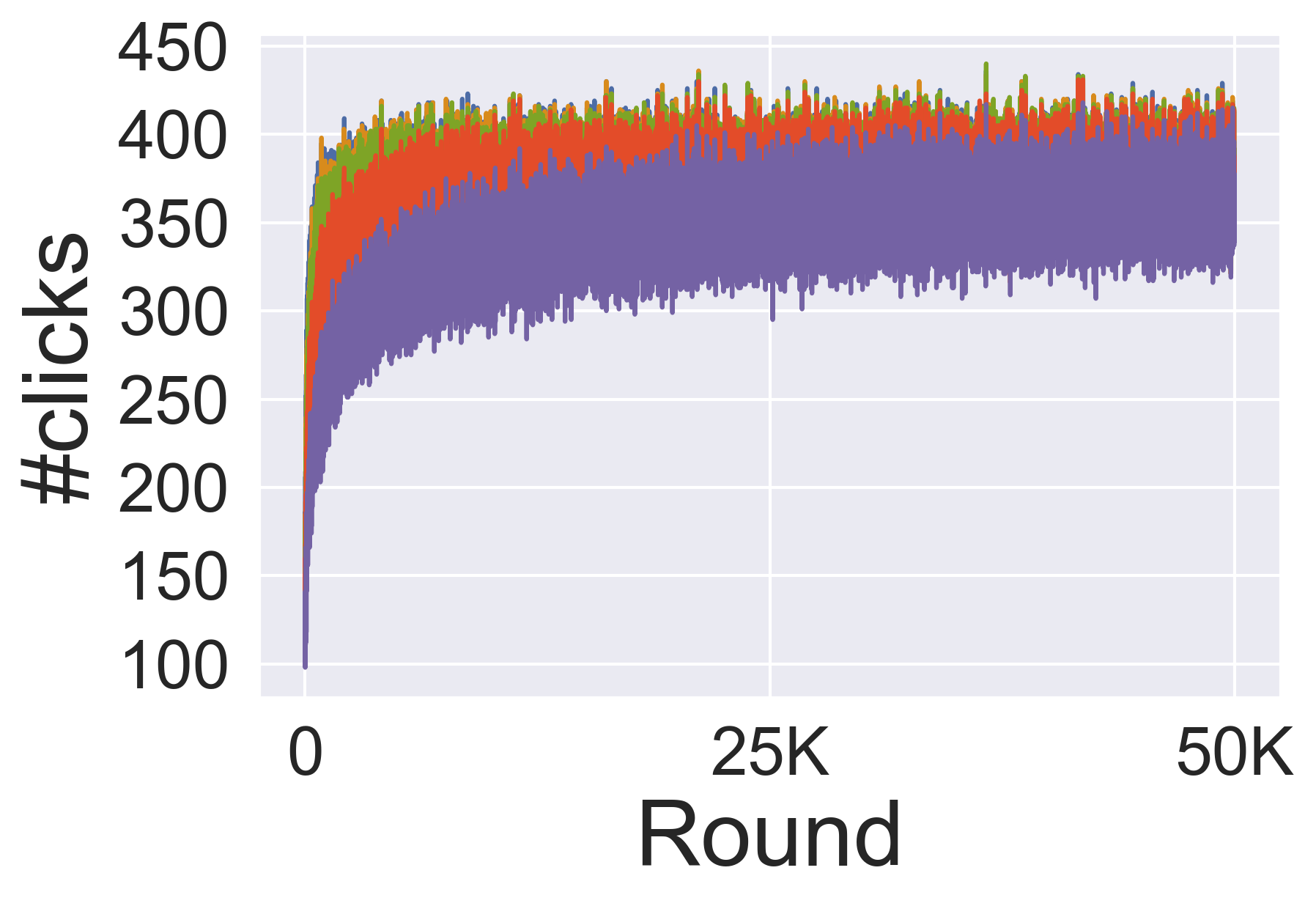}
      \caption{\#clicks per round}\label{fig:ml_alpha_cum_clicks}
    \end{subfigure}\hfill
    \begin{subfigure}{0.33\linewidth}
      \includegraphics[width=\linewidth]{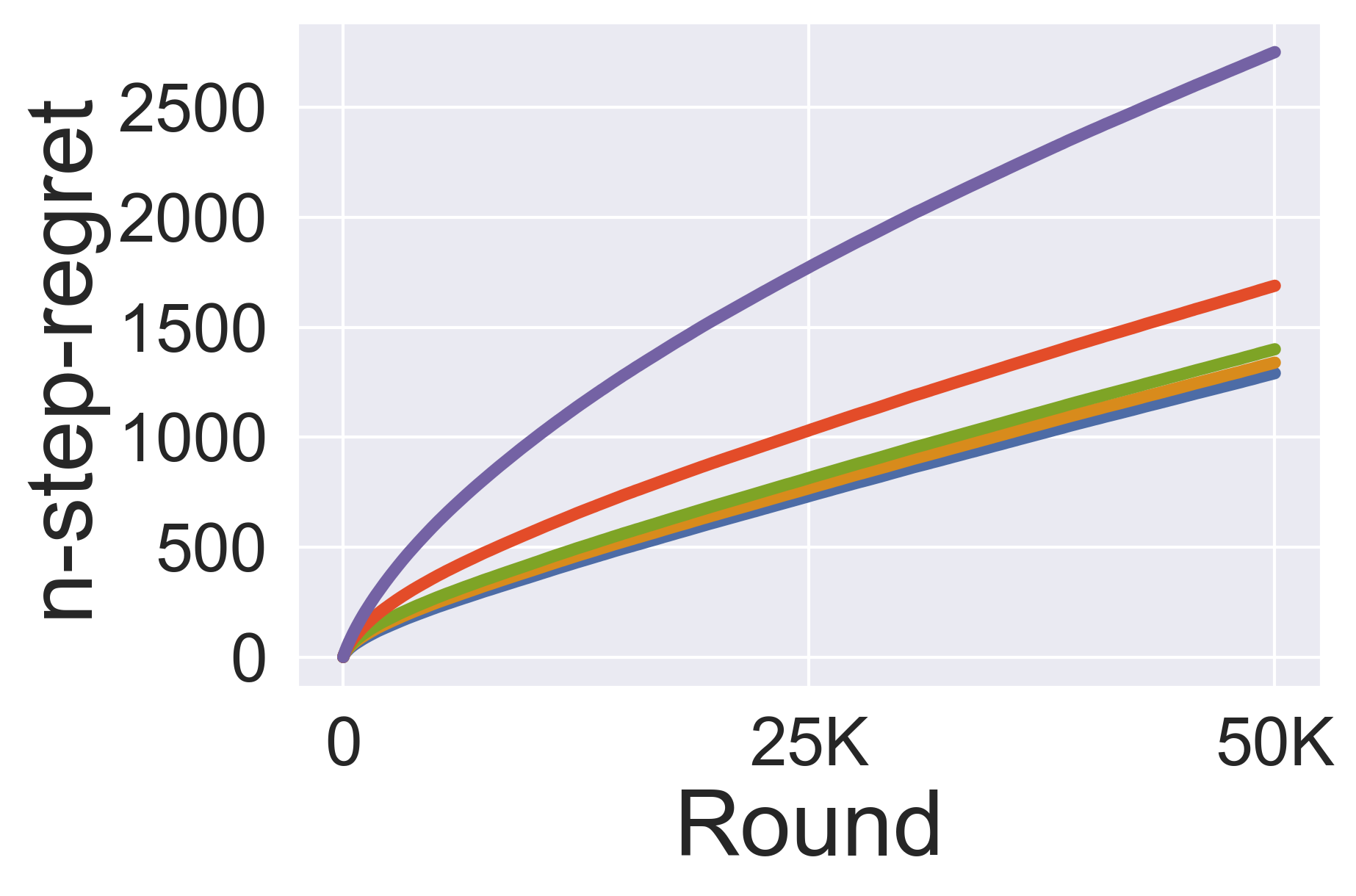}
      \caption{$n$-step-regret}\label{fig:ml_alpha_regret}
    \end{subfigure}
    \begin{subfigure}{0.33\linewidth}
      \includegraphics[width=\linewidth]{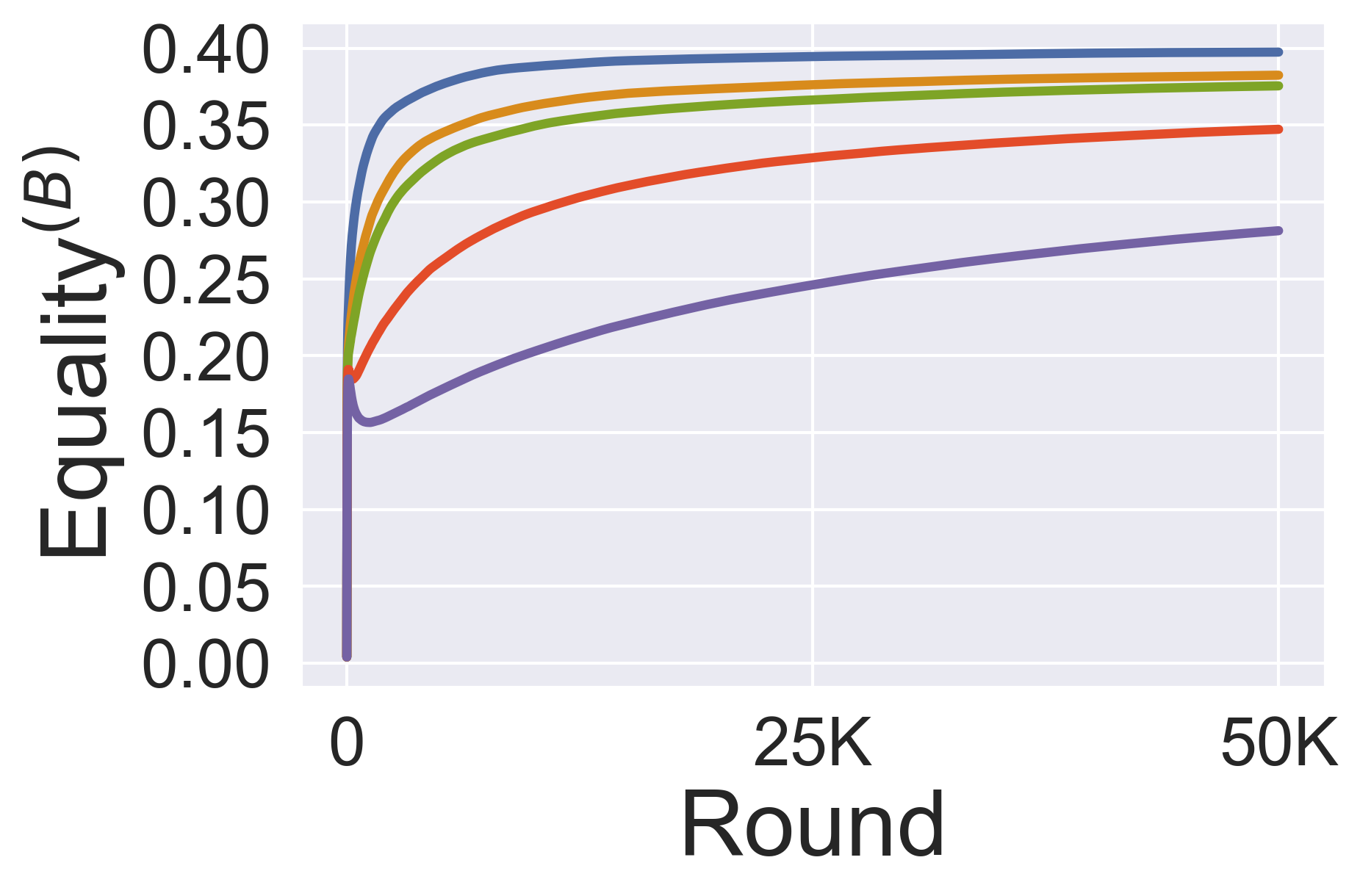}
      \caption{Equality$^{(B)}$ per round}\label{fig:ml_alpha_cum_equality_b}
    \end{subfigure}

    \medskip
    \begin{subfigure}{0.33\linewidth}
      \includegraphics[width=\linewidth]{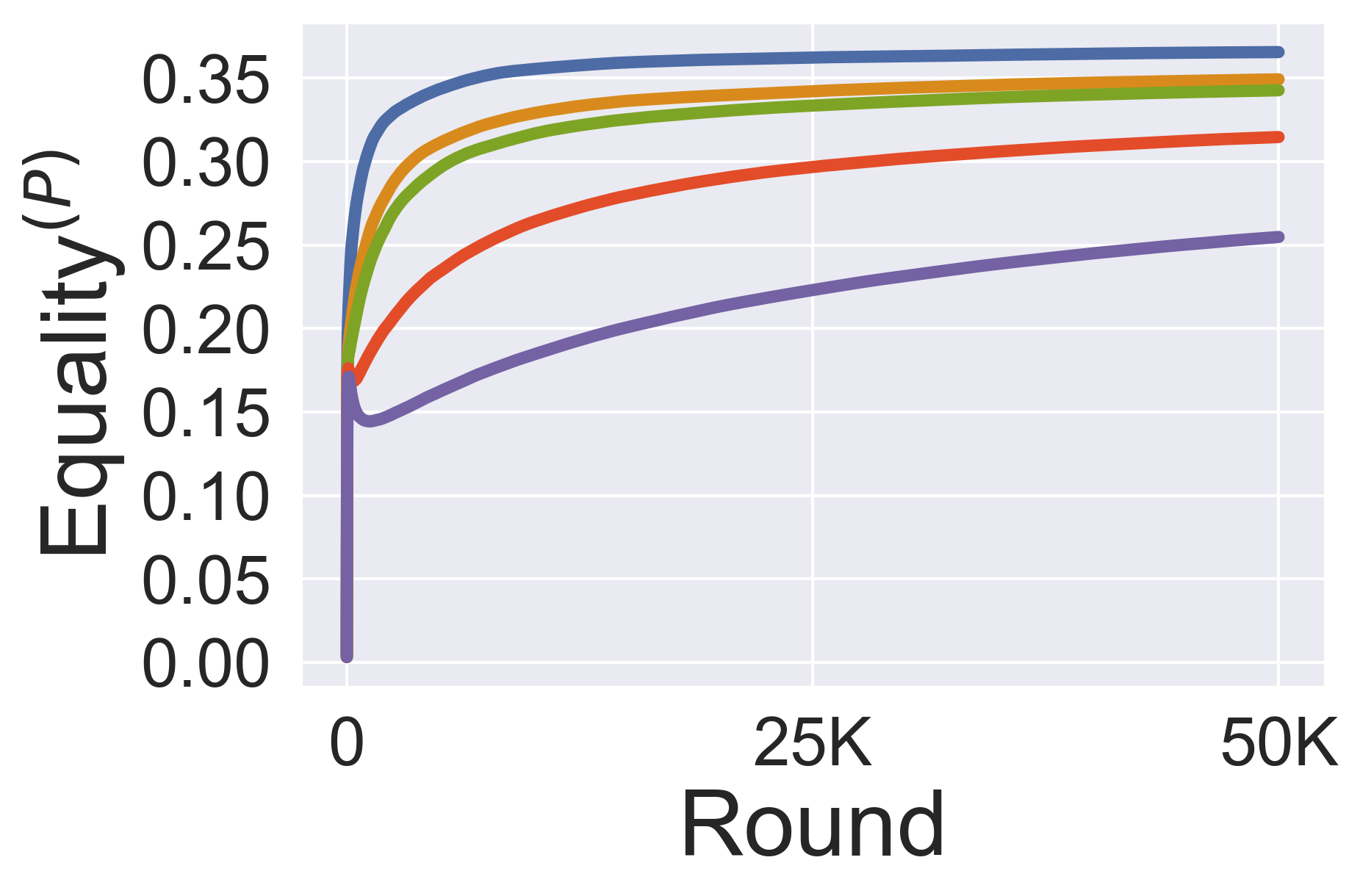}
      \caption{Equality$^{(P)}$ per round}\label{fig:ml_alpha_cum_equality_p}
    \end{subfigure}\hfill
    \begin{subfigure}{0.33\linewidth}
      \includegraphics[width=\linewidth]{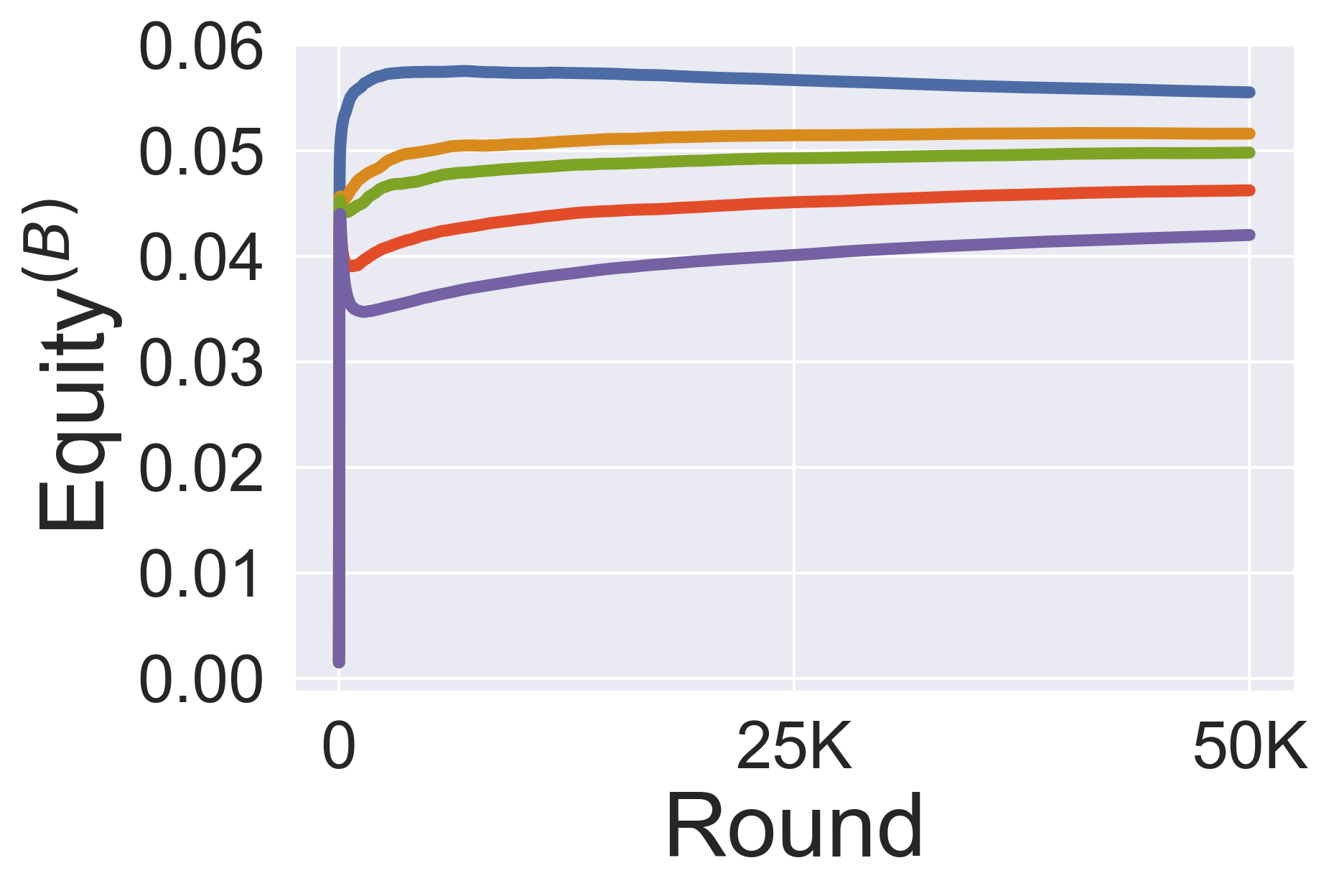}
      \caption{Equity$^{(B)}$ per round}\label{fig:ml_alpha_cum_equity_b}
    \end{subfigure}
    \begin{subfigure}{0.33\linewidth}
      \includegraphics[width=\linewidth]{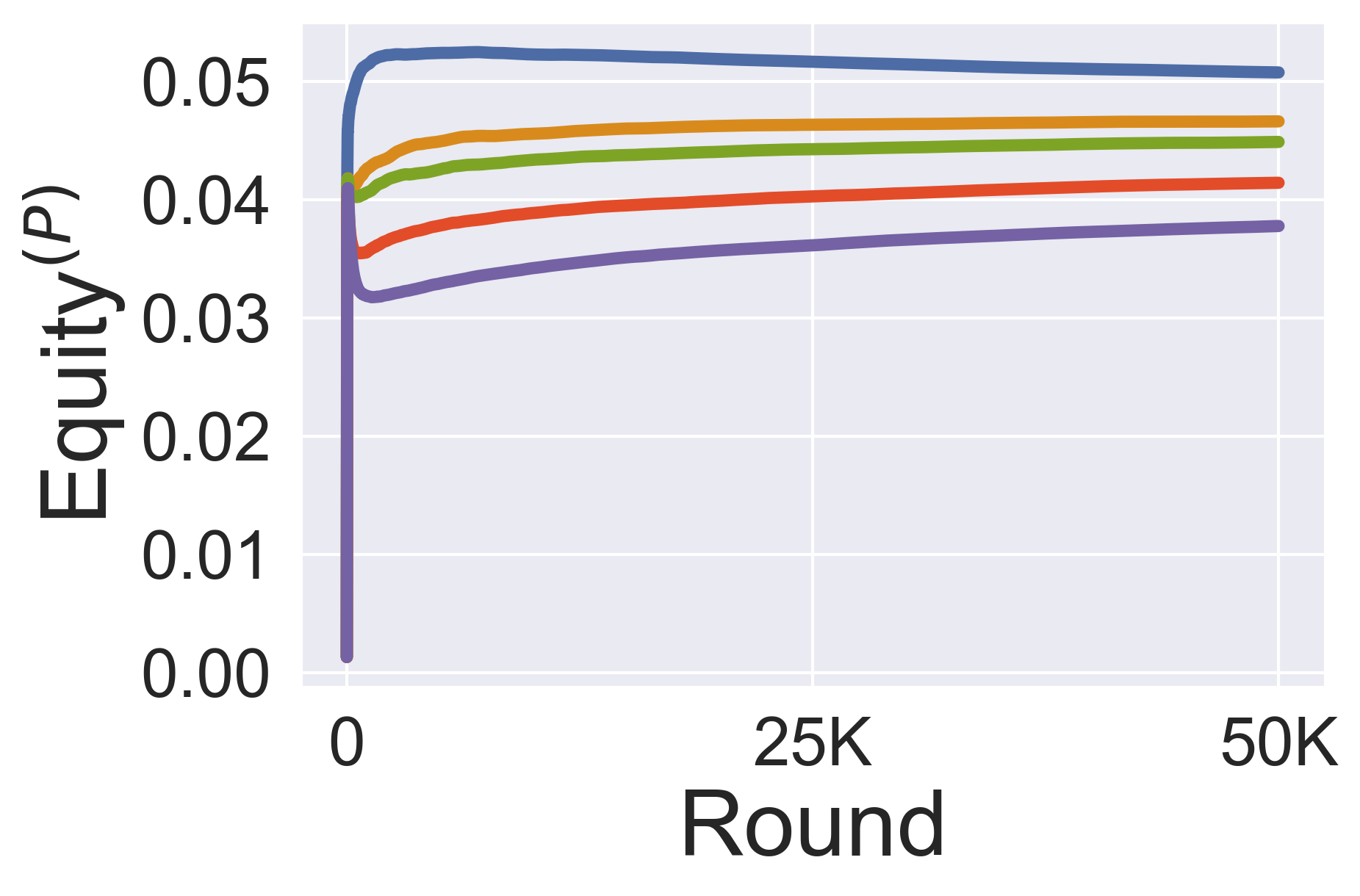}
      \caption{Equity$^{P}$ per round}\label{fig:ml_alpha_cum_equity_p}
    \end{subfigure}
  \end{minipage}
  \vspace{-8pt}
    \caption{The effect of varying the degree of exploration with $\alpha \in \{0.25,0.75,1,2,5\}$ on the performance of \algname{LinUCB} in terms of clicks and exposure bias on MovieLens dataset for $d=10$ and $K=10$. Left plot shows the average exploration across all users at each round, exploration is computed using the second term in Eq. \ref{eq_ucb}. Right plots: (a) number of observed clicks in each round, b) $n$-step-regret as in Eq. \ref{regret}, c-f) fairness metrics computed on accumulated exposure values at each round.}
    \label{fig:ml_alpha}
\end{figure*}

\subsection{Simulation and experimental setup}\label{sec_exp_setup}

The evaluation of interactive recommendation algorithms is usually done using off-policy evaluation approaches \cite{zhan2021off,wang2017optimal}. However, because in our problem the action space is too large (i.e. exponential in $K$), we utilize a simulated interaction environment for our evaluation where the simulator is built based on offline datasets. This is the evaluation setup used in similar research involving cascading bandits \cite{zong2016cascading,hiranandani2020cascading,li2020cascading}, which we also follow for our experiments.

We randomly divide the user profiles into training and test sets, with 50\% assigned to each. The training set is used to derive known variables and generate recommendations for users. The test set is used to model user feedback on recommendations and generate the optimal recommendation list to evaluate model performance.

To define the merits of the items, we adhere to definitions established in the existing literature \cite{balagopalan2023role,raj2022measuring,liu2024measuring}, where the relevancy of an item to users serves as its merit. To determine this, we employ a matrix factorization model on the user-item interaction data to learn the embeddings for users and items. Subsequently, we compute the relevance score between each user-item pair by taking the dot product of their embeddings. Finally, we compute the average relevance score across all users for each item, representing its merit.

We performed experiments with different dimensions of item embeddings $d \in \{10,20\}$ and different recommendation sizes $K \in \{5,10\}$. We tune each bandit algorithm with varying degrees of exploration $\alpha \in \{0.25,0.75,1,2,5\}$. Our \algname{EALinUCB} involves a hyperparameter, the penalization coefficient $\gamma$, for which we performed a sensitivity analysis with values $\gamma \in \{0.001,0.005,0.01,0.05,0.1,0.2\}$. We set the patient parameter $\beta$ involved in the weight functions to $\beta=0.05$ for Linear and $\beta=0.9$ for RBP. The experiments were carried out over $n=50,000$ rounds.

\section{Results}

In this section, we provide evidence and observations from our experimental results to address our three research questions\footnote{We report partial results in this paper. The full results are available at \url{https://github.com/masoudmansoury/ealinucb}.}.

\subsection{(RQ1) The effect of exploration degree on exposure bias in \algname{LinUCB}}

\textbf{RQ1} explores the relationship between the degree of exploration in the UCB item selection strategy, controlled by the hyperparameter $\alpha$ (Eq. \ref{eq_ucb}), and its impact on exposure bias and performance of \algname{LinUCB}. We examine this relationship using experimental results obtained from the MovieLens dataset for $d=10$ and $K=10$.

Figure \ref{fig:ml_alpha} (left) shows the average degree of exploration among all users in each round for varying values of $\alpha$. Here, exploration refers to the second term in Eq. \ref{eq_ucb}, which is computed for each recommended item for each user in each round. In particular, the average exploration value is derived by averaging exploration values across all recommended items for each user, and then averaging these values across all users at each round.

Several patterns emerge from Figure \ref{fig:ml_alpha} (left). Increasing the value of $\alpha$ leads to a higher degree of exploration. Secondly, the degree of exploration rapidly decreases after several rounds. For example, with $\alpha=2$, exploration substantially decreases after approximately 1000 rounds, reaching 0 around 10,000 rounds. This behavior aligns with the exploration/exploitation trade-off in bandit algorithms \cite{barraza2017exploration,cao2024does}: as the algorithm accumulates more information over time, exploration decreases and exploitation increases.

\begin{table*}
	\centering
        \caption{Performance of our \algname{EALinUCB} with three different weight functions on MovieLens and Yahoo Music datasets for $d=10$ and $K=5$. For all metrics, higher value is more desired.
        $\dag$ indicates that the result is significant with $p < 0.01$.} %
        \vspace{-12pt}
		\label{tab:d10k5}%
        \begin{subfigure}{\linewidth}
            \centering
			\begin{tabular}{l l l l l l l l l l l l}
			\toprule
            \multirow{2}[1]{*}{Method} & \multirow{2}[1]{*}{$\mathcal{F}$} & \multicolumn{5}{c}{ML} & \multicolumn{5}{c}{Yahoo Music} \\
            \cmidrule(r){3-7} \cmidrule(r){8-12}
		& & $\overline{clicks}$ & Equality$^B$ & Equality$^P$ & Equity$^B$ & Equity$^P$ & $\overline{clicks}$ & Equality$^B$ & Equality$^P$ & Equity$^B$ & Equity$^P$ \\
            \midrule
            \algname{LinUCB}&-&\textbf{0.2166}&0.329&0.3081&0.0506&0.0476&0.1802&0.3602&0.3316&0.3559&0.3278\\
            \algname{EARSLinUCB} & -&0.2165&0.3257&0.3257&0.0495&0.0495&0.1807&0.3591&0.3591&0.3556&0.3556\\
            \algname{FRMLinUCB}         & -     &0.2005&0.3334&0.4586&0.0504&0.0505&0.1721&0.3514&0.42&0.3514&0.3501\\
            \algname{EALinUCB} (ours)   & Log   &0.2105&0.3799&\textbf{0.524}$^\dag$&\textbf{0.055}&\textbf{0.0546}&0.1772&\textbf{0.3662}&0.5396&\textbf{0.3599}&\textbf{0.3641}\\
            \algname{EALinUCB} (ours)   & RBP   &\textbf{0.2166}&0.329&0.472&0.0506&0.0515&\textbf{0.1814}&0.365&0.5812&0.358&0.362\\
            \algname{EALinUCB} (ours)   & Linear&0.2069&\textbf{0.392}$^\dag$&0.5142&0.0545&0.0535&0.1706&0.3661&\textbf{0.5879}$^\dag$&0.3563&0.3582\\
              \bottomrule
           \end{tabular}
    \end{subfigure}
\end{table*}%

\begin{table*}
	\centering
        \vspace{-5pt}
        \caption{Performance of our \algname{EALinUCB} with three different weight functions on MovieLens and Yahoo Music datasets for $d=10$ and $K=10$. For all metrics, higher value is more desired.
        $\dag$ indicates that the result is significant with $p < 0.01$.} %
        \vspace{-12pt}
		\label{tab:d10k10}%
        \begin{subfigure}{\linewidth}
            \centering
			\begin{tabular}{l l l l l l l l l l l l}
			\toprule
            \multirow{2}[1]{*}{Method} & \multirow{2}[1]{*}{$\mathcal{F}$} & \multicolumn{5}{c}{ML} & \multicolumn{5}{c}{Yahoo} \\
            \cmidrule(r){3-7} \cmidrule(r){8-12}
		& & $\overline{clicks}$ & Equality$^B$ & Equality$^P$ & Equity$^B$ & Equity$^P$ & $\overline{clicks}$ & Equality$^B$ & Equality$^P$ & Equity$^B$ & Equity$^P$ \\
            \midrule
            \algname{LinUCB}&-&\textbf{0.3722}&0.3974&0.3656&0.0555&0.0507&0.3154&0.4469&0.404&0.4429&0.4002\\
            \algname{EARSLinUCB}&-&0.3721&0.3974&0.3848&0.0562&0.0541&0.3157&0.4467&0.4467&0.4429&0.4429\\
            \algname{FRMLinUCB}&-&0.3574&0.4029&0.4157&0.0572&0.055&0.3085&0.4517&0.4596&0.4517&0.4498\\
            \algname{EALinUCB} (ours)&Log&0.3605&0.494&\textbf{0.514}$^\dag$&\textbf{0.065}$^\dag$&\textbf{0.065}$^\dag$&0.3073&\textbf{0.495}$^\dag$&0.5174&\textbf{0.485}$^\dag$&\textbf{0.473}$^\dag$\\
            \algname{EALinUCB} (ours)&RBP&\textbf{0.3722}&0.4074&0.434&0.0555&0.0557&\textbf{0.3171}&0.4489&\textbf{0.553}$^\dag$&0.4614&0.4585\\
            \algname{EALinUCB} (ours)&Linear&0.3585&\textbf{0.498}$^\dag$&0.5062&0.0601&0.0604&0.3008&0.4552&0.5312&0.461&0.4542\\
              \bottomrule
           \end{tabular}
    \end{subfigure}
\end{table*}%

Figures \ref{fig:ml_alpha_cum_clicks}-\ref{fig:ml_alpha_cum_equity_p} depict the performance of on \algname{LinUCB}' with varying $\alpha$ values. It can be observed that exploration negatively affects the accuracy of the model. Looking at Figure \ref{fig:ml_alpha_cum_clicks}, lower clicks are observed for \algname{LinUCB} with higher $\alpha$ values (e.g., $\alpha=5$) compared to lower values (e.g., $\alpha=1$). Furthermore, in Figure \ref{fig:ml_alpha_regret}, \algname{LinUCB} consistently exhibits a better $n$ step-regret with lower $\alpha$ values (e.g., $\alpha=1$) compared to higher values (e.g., $\alpha=5$).

It is also evident from Figures \ref{fig:ml_alpha_cum_equality_b}-\ref{fig:ml_alpha_cum_equity_p} that only during the exploration phase is exposure fairness improving. However, after the model stops exploring the item space, the exposure bias does not decrease any further. Although this is the normal behavior of bandit algorithms, our aim is to achieve a higher degree of fairness before the algorithm stops the exploration. In addition, the plots show that for various values of $\alpha$, the degree to which the exposure bias decreases is different. Surprisingly, a higher $\alpha$ value does not result in a higher exposure fairness for items. Hence, this confirms the necessity of an intervention in \algname{LinUCB}, as its built-in exploration component does not adequately mitigate exposure bias. These results are consistent with findings in \cite{mansoury2022exposure,mansoury2021unbiased}.
Since \algname{LinUCB} with $\alpha=0.25$ yields the best performance across all the metrics, for the rest of the analysis in this paper, we set $\alpha=0.25$.

\subsection{(RQ2) Comparison to baselines}\label{sec_rq2}

\begin{figure}
  \centering
      \includegraphics[width=\linewidth]{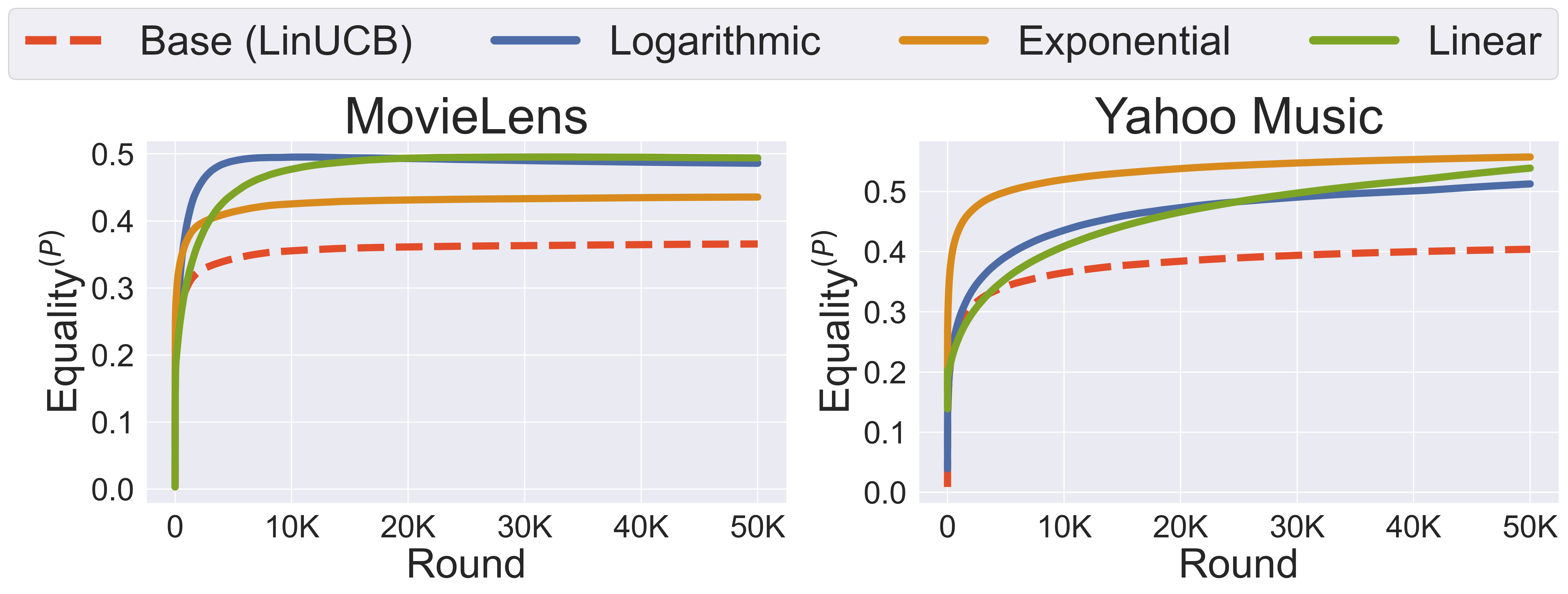}
    \vspace{-22pt}
    \caption{Comparison of \algname{LinUCB} and \algname{EALinUCB} with three weight functions in terms of Equality$^{(P)}$ per round for $d=10$, $K=10$, $\alpha=0.25$, and $\gamma=0$. At each round $t$, Equality$^{(P)}$ is computed over the accumulated exposure up to round $t$.}
    \vspace{-8pt}
    \label{fig:exposure_p_ea}
\end{figure}

To address \textbf{RQ2}, we compare the performance of our \algname{EALinUCB} with other baselines using different exposure bias metrics. We set $\alpha=0.25$ and $\gamma=0$ for all experiments. Tables \ref{tab:d10k5} and \ref{tab:d10k10} present the results for $d=10$ and $K \in \{5,10\}$. These results indicate that \algname{EALinUCB} outperforms other algorithms consistently across all exposure bias metrics. This improvement is often significant, demonstrating its effectiveness in mitigating exposure bias. Figure \ref{fig:exposure_p_ea} compares the Equality$^{(P)}$ per round between \algname{EALinUCB} and \algname{LinUCB} for $d=10$, $K=10$, $\alpha=0.25$, and $\gamma=0$. The plot reveals that \algname{EALinUCB} significantly enhances the exposure fairness in the long run, particularly with logarithmic and linear weight functions.

\begin{figure*}
  \centering
    \begin{subfigure}{1\linewidth}
      \includegraphics[width=\linewidth]{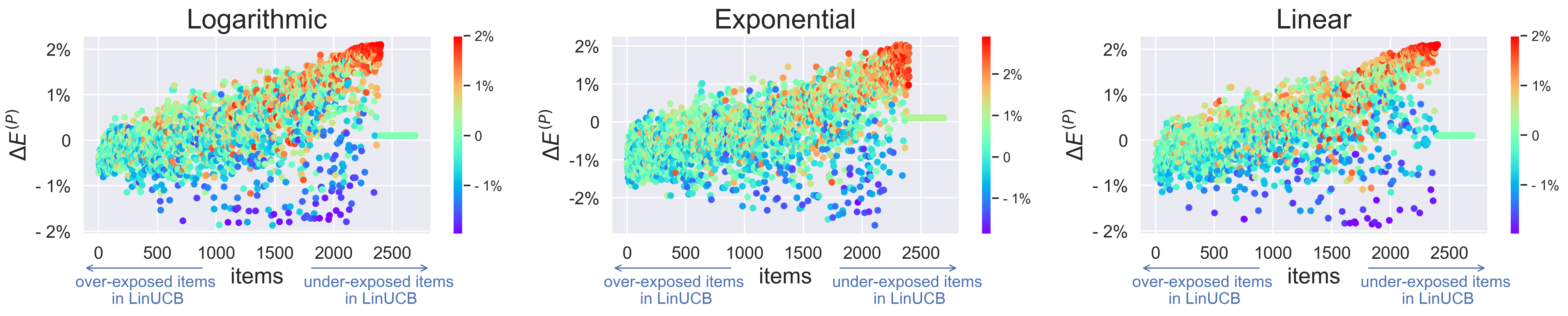}
      \vspace{-17pt}
      \caption{MovieLens}\label{fig:ml_exposure_b}
    \end{subfigure}\hfill
    \begin{subfigure}{1\linewidth}
      \includegraphics[width=\linewidth]{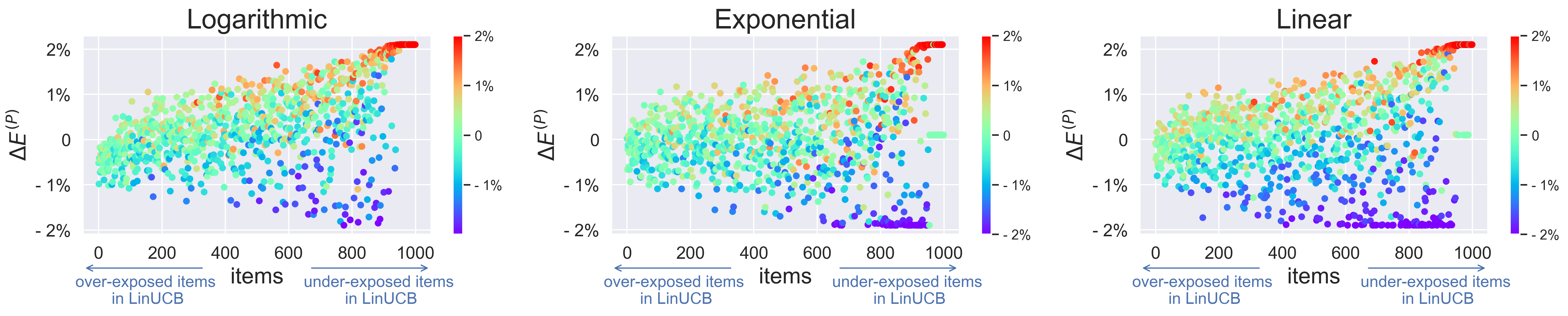}
      \vspace{-17pt}
      \caption{Yahoo Music}\label{fig:r2y_exposure_b}
    \end{subfigure}\hfill
    \vspace{-10pt}
    \caption{Exposure analysis of our \algname{EALinUCB} with three different weight functions for $d=10$ and $K=10$. Colorbar shows the percentage increase/decrease in $\overline{clicks}$. Items are sorted based on their exposure ($E^{(P)}$) by \algname{LinUCB} in descending order from left to right where items in the left-side are the over-exposure ones and items in the right-side are under-exposed ones.}
    \label{fig:exposure_b}
\end{figure*}

Improving exposure fairness involves balancing the exposure for items by downgrading over-exposed items that are often ignored by users and promoting under-exposed items that are clicked more often. 
To examine the ability of \algname{EALinUCB} to balance exposure compared to \algname{LinUCB}, we calculate the percentage change in exposure for each item assigned by \algname{EALinUCB} compared to \algname{LinUCB} as:
\vspace{-3pt}
\begin{equation}\label{eq_delta_exposure}
    \vspace{-5pt}
    \Delta E(i)=\frac{E_{EALinUCB}(i)-E_{LinUCB}(i)}{\frac{E_{EALinUCB}(i)+E_{LinUCB}(i)}{2}} \times 100
\end{equation}
\noindent where $E_{EALinUCB}(i)$ and $E_{LinUCB}(i)$ are the exposure given to item $i$ by \algname{EALinUCB} and \algname{LinUCB}, respectively. Analogously, we compute the percentage change in $\overline{clicks}$ observed in each element by \algname{EALinUCB} and \algname{LinUCB}. Figure \ref{fig:exposure_b} shows how our \algname{EALinUCB} assigns exposure to each item compared to \algname{LinUCB} for $E^{(P)}$ exposure definition. 
The x-axis displays items sorted by $E^{(P)}$ by \algname{LinUCB} in descending order, the y-axis shows $\Delta E^{(P)}$ computed by Eq. \ref{eq_delta_exposure}. The color bar also shows $\Delta \overline{clicks}$.

\begin{figure*}
  \centering
      \includegraphics[width=0.95\linewidth]{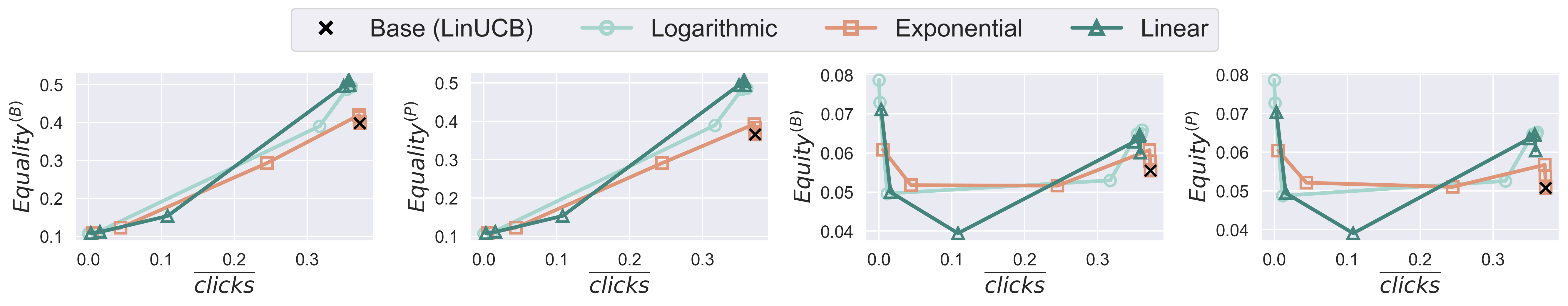}
      \vspace{-15pt}
    \caption{Performance of our \algname{EALinUCB} with three different weight functions in terms of $\overline{clicks}$ and fairness metrics for varying $\gamma \in \{0,0.001,0.005,0.01,0.05,0.1,0.2\}$ on MovieLens dataset for $d=10$ and $K=10$. The cross shows the performance of \algname{LinUCB}.} 
    \label{fig:sensitivity}
\end{figure*}

Figure \ref{fig:exposure_b} indicates that \algname{EALinUCB} promotes under-exposed items in \algname{LinUCB} while downgrading over-exposed ones, effectively balancing exposure for different items. In addition, the red points on the upper right indicate that \algname{EALinUCB} predominantly promotes relevant items, as most of the promoted items also receive more clicks. These patterns are consistent across all datasets and weight functions. Similar patterns are observed for other exposure notions (reported in this \href{https://github.com/masoudmansoury/ealinucb}{\textcolor{blue}{\underline{website}}}).

\subsection{(RQ3) The impact of varying penalization degree ($\gamma$)}\label{sec_rq3}

Our exposure-aware reward model, as defined in Eq. \ref{eq_eareward}, involves a hyperparameter $\gamma$ that regulates the extent of penalization for unclicked items. To explore the sensitivity of \algname{EALinUCB} to different values of $\gamma$, we conducted a sensitivity analysis on the MovieLens dataset for $d=10$ and $K=10$, varying $\gamma$ within the range $\{0,0.001,0.005,0.01,0.05,0.1,0.2\}$. Figure \ref{fig:sensitivity} presents the results of this analysis. Each plot corresponds to a specific exposure bias metric, with the $\overline{Clicks}$ values plotted on the x axis and the metric values on the y-axis. The points on the right side of the plots represent the results for $\gamma=0$, while the points on the left show the results for $\gamma=0.2$. The crosses represent the performance of \algname{LinUCB} as the base algorithm.

The results show that with a proper choice of $\gamma$ value, the penalization term can have a positive impact on mitigating exposure bias. For example, in all plots corresponding to the exponential weight function, increasing the value of $\gamma$ from 0 to 0.005 leads to improvements in all exposure metrics. However, further increasing the value of this hyperparameter results in a decrease in performance. When $\gamma=0.2$, for example, $\overline{clicks}$ approaches 0, indicating deteriorating performance, along with reductions in exposure metrics, which means increased exposure bias.

The observed trend can be attributed to the dominance of the penalization term in the learning process, especially with higher values of $\gamma$. When $\gamma$ is large, the algorithm predominantly learns negative preferences due to the abundance of unclicked items compared to clicked items. Consequently, the performance of \algname{EALinUCB} declines. Hence, careful tuning of this hyperparameter is essential to optimize the algorithm performance.

\section{Conclusion and Future Work}

In this paper, we studied the problem of exposure bias in linear cascading bandits. Although these algorithms partially mitigate exposure bias during the initial exploration phase, we show their limitations in balancing item exposure over the recommendation lifecycle. To improve exposure fairness throughout the recommendation process, we introduced an \textit{exposure-aware reward model} and integrated it into the linear cascading bandit. This model leverages user feedback and item position in the recommendation list to reward clicked items and penalize unclicked ones. Our extensive experiments demonstrated the effectiveness of the proposed exposure-aware reward model in mitigating exposure bias while preserving recommendation accuracy. Additionally, we theoretically derived a gap-free bound on the $n$-step-regret for our exposure-aware cascading bandit.
In future work, we plan to extend our analysis to other cascading bandits~\cite{hiranandani2020cascading,li2020cascading} as well as broader classes of bandit algorithms like those based on Thompson Sampling.

\begin{acks}
This project was funded by Elsevier’s Discovery Lab.
\end{acks}

\newpage
\balance
\bibliographystyle{ACM-Reference-Format}
\bibliography{ref}

\end{document}